\documentclass[fleqn,usenatbib]{mnras}

\usepackage{newtxtext,newtxmath}
\usepackage{soul}
\usepackage{xcolor}
\sethlcolor{yellow}
\usepackage[T1]{fontenc}

\DeclareRobustCommand{\VAN}[3]{#2}
\let\VANthebibliography\thebibliography
\def\thebibliography{\DeclareRobustCommand{\VAN}[3]{##3}\VANthebibliography}

\usepackage{graphicx}
\usepackage{caption,subcaption}
\defcitealias{Somawanshi}{Paper I}
\title[Chemically identified dwarf using MUSE]{Chemical signature reveals co-spatial dwarf satellite of an edge-on disc galaxy with MUSE}

\author[Devang Somawanshi et al.]
{Devang Somawanshi,$^{1}$\thanks{E-mail: devangsomawanshi11@gmail.com}
Souradeep Bhattacharya,$^{2,3}$\thanks{E-mail: s.bhattacharya3@herts.ac.uk}
Manish Kataria,$^{3}$ 
Preetish K. Mishra,$^{4}$ and 
\newauthor
Chiaki Kobayashi$^{2}$ \\
{$^{1}$Department of Physics, Indian Institute of Science Education and Research Bhopal, Bhopal Bypass Road, Bhauri, Bhopal 462 066, Madhya Pradesh, India}\\
{$^{2}$Centre for Astrophysics Research, Department of Physics, Astronomy and Mathematics, University of Hertfordshire, Hatfield, AL10 9AB, UK}\\
{$^{3}$Inter University Centre for Astronomy and Astrophysics, Ganeshkhind, Post Bag 4, Pune 411007, India}\\
{$^{4}$School of Physics, Korea Institute for Advanced Study, 85 Hoegiro, Dongdaemun-gu, Seoul 02455, Republic of Korea}\\
}

\date{Accepted XXX. Received YYY; in original form ZZZ}

\pubyear{2025}

\begin{document}
\label{firstpage}
\pagerange{\pageref{firstpage}--\pageref{lastpage}}
\maketitle

\begin{abstract}
Integral field unit (IFU) spectroscopic observations of resolved galaxies provide an optimal experimental setting for determination of stellar population properties, in particular - age, metallicity and $\alpha$-enhancement, which are key to understanding evolution of galaxies across diverse physical environments. We determine these properties for the edge-on disc galaxy IC~1553, through stellar population models fitted to MUSE IFU observations. From our determined spatial distributions of metallicity and [$\alpha$/Fe], we serendipitiously identify the unique chemical signature of a dwarf galaxy that is co-spatial with the luminous disc of IC~1553. The dwarf galaxy is characterized by the presence of higher [$\alpha$/Fe] and metal-poor stellar populations relative to the disc of IC~1553. The identified dwarf is dynamically cold from its determined kinematics, consistent with being a satellite of IC 1553. From modeling the \textit{Spitzer} IRAC 3.6~$\mu m$ image of IC~1553, we confirmed the presence of the dwarf galaxy and calculated its stellar mass to be $\sim1.28\times 10^{9} \rm~M_{\odot}$. This is the first such identification of a dwarf galaxy from its unique chemical signature in such integrated light IFU observations, even though its hidden by the luminous body of its massive host.
\end{abstract}


% Introduction to last section, including all figures (not acknowledgements and beyond) needs to be within 5 pages for an MNRAS letter

% %Souradeep-I have used clearpage to deleneate the body of the paper

%%%%%%%%%%%%%%%%%%%%%%%%%%%%%%%%%%%%%%%%%%%%%%%%%%
%%%%%%%%%%%%%%%%% BODY OF PAPER %%%%%%%%%%%%%%%%%%

\section{Introduction}
\label{sect: intro}
Numerous dwarf galaxies, typically with baryonic masses of approximately $10^9\text{M}_\odot$ or less, have begun to be identified routinely in photometric sky surveys  (e.g. Palomar Sky Survey: \citealt{Schombert97}; Sloan Digital Sky Survey: \citealt{Brinchmann04}; Hyper Suprime-Cam Subaru Strategic Program: \citealt{Aihara22}) as well as in dedicated photometric searches for satellite dwarfs, including faint ones, near massive galaxies (e.g. \citealt{Venhola18}; see review by \citealt{Karachentsev19} for additional references). Given the high luminosity of such massive galaxies relative to their dwarf satellites, detecting dwarf satellites that are co-spatial with their massive companion presents considerable difficulty, especially when both the massive galaxy and the dwarf are star-forming. For nearby galaxy groups like M81 and Cen~A, such dwarf galaxies may be identified from deep imaging surveys of resolved stars, if their stellar populations distribute differently from that of their massive host galaxies in color-magnitude-diagrams \citep[e.g.][]{Chiboucas13, Crnojevic16}. But such deep photometric surveys resolving individual stars are more difficult for more distant galaxies. If the spatial distribution of line-of-sight velocities (LOSV) and LOSV dispersions (LOSVD) are available for the massive galaxy, then kinematically discordant co-spatial dwarfs may be identified even if their stellar populations are not resolved \citep[e.g.][]{hartke18}. However identifying a co-spatial dwarf is even more of a challenge when the dwarf galaxy does not stand out clearly is the LOSV and LOSVD maps.

Dwarf galaxies, including those that are co-spatial with their massive star-forming companions, are however expected to have stellar population properties distinct from massive star-forming spiral galaxies. Indeed, they are composed of stellar populations that, on average, have lower metallicity, following the mass metallicity relation \citep{2011Kirby,Curti20}. The elemental abundance ratios are also different; [$\alpha$/Fe] of low-mass galaxies are lower than those of massive galaxies \citep{Thomas10,Spolaor10}. At a given metallicity [$\alpha$/Fe] of such stars in low-mass galaxies are also lower than the MW halo, but higher than MW disk \citep{Tolstoy09}. These are the result of the interplay of different chemical yields from Core-collapse (CCSNe) and Type Ia (SNe Ia) Supernovae given the difference in star-formation histories between dwarfs and spirals \citep{Kobayashi20b,Kobayashi20}.

Stellar population model fits to integrated galaxy spectra can constrain galaxy stellar population properties \citep[e.g.][]{McDermid15}. Deep integral field unit (IFU) spectroscopy observations of galaxies, such as with MUSE at the Very Large Telescope \citep{2010Bacon}, in conjunction with such stellar population model fitting have enabled the mapping of stellar population properties like age, metallicity and [$\alpha$/Fe] \citep[e.g.][]{Comeron2016,2016Kasparova, 2019PinnaA,2019PinnaB,2021Scott}. Of particular interest are edge-on star-forming disc galaxies, where distinct stellar populations properties (mainly [$\alpha$/Fe]) have been determined for their thin and thick discs \citep[e.g.][hereafter Paper I]{2021Scott, Somawanshi}.

In this work, we report the serendipitous identification of a dwarf satellite galaxy, co-spatial with an edge-on disc galaxy IC~1553. Based on stellar population model fitting to MUSE IFU measurments, we find that the dwarf galaxy shows a distinct chemical signature compared to its massive companion. Its identification is supported with \textit{Spitzer space telescope} imaging. We describe the basic characteristics of IC~1553 along with a discussion of the MUSE and \textit{Spitzer} data in Section~\ref{sect: obs}. The methodologies employed for stellar population analysis resulting in the identification of the dwarf galaxy, as well as its kinematics, are presented in Section~\ref{sect: analysis}. The IC~1553 \textit{Spitzer} image modeling  and mass estimation of the dwarf galaxy are outlined in Section~\ref{sect: Imaging_analysis}. Finally, Section~\ref{sect: discussion_and_conclusion} encompasses a discussion and conclusions regarding the nature of the dwarf galaxy associated with IC~1553.%{\color{red} [we should also mention the cosmology here, like 737, as we are quoting distances.] - Distance is determined by Tully and we use the reported distance. }

%This trend is also evident in dwarf spheroidal (dSph) satellites of the Milky Way, where the average [$\alpha$/Fe] declines from $\sim$+0.4 at [Fe/H] $\sim$ -2.5 to $\sim$0.0 at [Fe/H] $\sim$ -1.2 \citep{2011Kirby}. M31's dwarf spheroidal satellites also display a similar decrease in [$\alpha$/Fe] with increasing [Fe/H], showing no significant deviation from the trends observed in the Milky Way's dSph satellites \citep{2020Kirby}. Some globular clusters in the dwarf spheroidal galaxy KK 84, the closest satellite of the S0 galaxy NGC 3115, exhibit elevated $\alpha$-abundances paired with near-solar metallicities \citep{2008Puzia}. 
%Furthermore, using spectra from the UVES spectrograph, \citet{2003Sheraton} and \citet{2005Geisler} identified an [$\alpha$/Fe]-enhanced population within the Sculptor dwarf spheroidal galaxy (Scl dSph) at [Fe/H] $\sim$ -1.8. Similarly, the chemical composition analysis of 14 stars in the Sagittarius dwarf spheroidal galaxy (Sgr dSph), observed with the Keck Observatory, demonstrated an [$\alpha$/Fe] enhancement reaching up to [Fe/H] $\sim$ -1.0 \citep{2005McWilliam}, indicating a comparatively metal-rich population relative to the Scl dSph. Comparable findings were also reported for the Sagittarius dwarf spheroidal galaxy (Sgr dSph) as a result of the FLAMES-UVES-VIMOS survey \citep{2005Monaco,2007Sbordone}.

\begin{figure}
\includegraphics[width=\columnwidth]{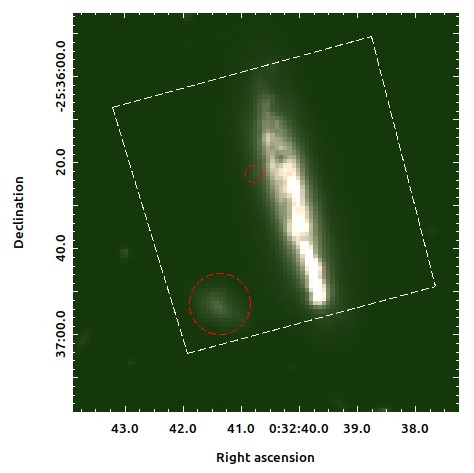}
\caption{Colour image of IC 1553 constructed from \textit{g,r,z}-band images from the Dark Energy Camera Legacy Survey (DECaLS DR9, \citealt{Dey19}). The MUSE field-of-view is marked in white. The two interloping objects that were removed from the analysed MUSE data cube are marked in red}.
\label{Fig: color}
\end{figure} 

%______________________________________%

\begin{figure}
    \includegraphics[width=0.5\textwidth]{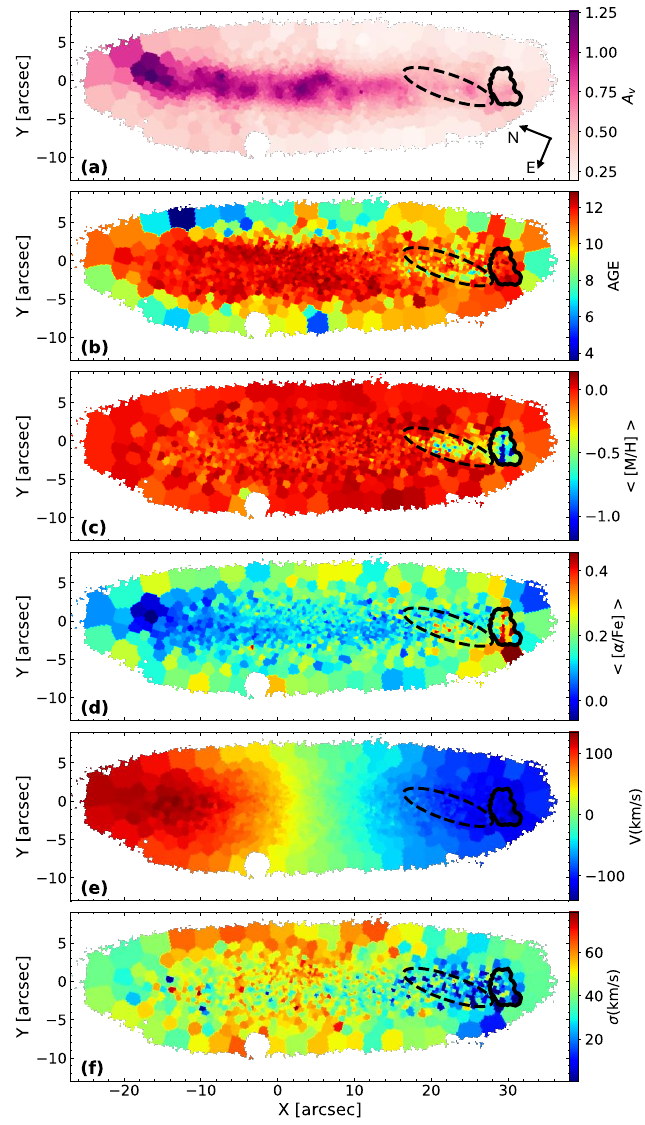}

    \caption{Spatial distribution of properties for each spatial bin in IC 1553 showing (a) line-of-sight extinction ($A_{v}$), (b) mass-weighted mean age, (c) mass-weighted mean metallicity <[M/H]>, (d) mass-weighted mean alpha abundance <[$\alpha$/Fe]>, (e) mean stellar velocity (V), and (f) mean stellar velocity dispersion ($\sigma$). The analyzed dwarf spatial bins are marked in solid lines. An additional related region of interest is marked with a dotted ellipse.}
    \label{Fig: Mean_values}
\end{figure}

\section{Observations of IC~1553}
\label{sect: obs}

IC~1553 (see Figure~\ref{Fig: color}) is an S-type galaxy located approximately 36.5 Mpc away \citep{Tully16}. With a stellar mass of $\sim 1.34 \times 10^{10} \rm~M_{\odot}$ \citep {Comerón2018}, it is comparable to that of the Milky Way (M$_{*}\sim 5.4 \times 10^{10} \rm~M_{\odot}$ \citealt {McMillan2017}). \citet{Lauberts1982} classified IC~1553 as a galaxy exhibiting extended stellar and gaseous material. It has also been classified as an emission-line galaxy exhibiting significant star formation activity \citep{RAUTIO}. It also exhibits notable dust extinction in its central region, with a prominent dust lane visible along its mid-plane \citep {Comerón2019}. IC~1553 was part of an initial subsample of 70 edge-on disc galaxies drawn from a larger sample of galaxies that constituted the S$^{4}$G survey \citep[\textit{Spitzer} Survey of Stellar Structure in Galaxies;][]{Sheth2010,Comerón2012}. These galaxies were mapped with \textit{Spitzer} IRAC channels 1 and 2 (3.6 and 4.5 $\mu m$). From amongst these edge-on discs, IC~1553 was included in the subsample of star-forming galaxies selected for IFU spectroscopic follow-up with MUSE by \citet{Comerón2019}. They identified its kinematically distinct thin and thick discs. Utilizing the same MUSE IFU data, \citet{RAUTIO} found an increasing radial velocity lag in the outskirts of the galaxy disc with H$\alpha$ kinematics, potentially due to the accretion of gas from a diffuse source.
 
IC 1553 forms part of our sample of edge-on disc galaxies with archival MUSE IFU observations by \citet{Comerón2019}, whose stellar population properties (particularly age, metallicity, and [$\alpha$/Fe]) we are studying to understand the formation and evolution of disc galaxies (Somawanshi et al. in prep). The stellar population analysis for a pilot galaxy, ESO~544-27, has been presented in \citetalias{Somawanshi}. 

%\subsection{MUSE IFU data}
%\label{sect: IFU}

Spatial sampling of MUSE is $0.2''$ with a $1' \times 1'$ field-of-view and a wavelength range from 4750--9351~\AA~ \citep{2010Bacon} and with a median spectral resolution of 2.5~\AA. IC 1553 was observed in four dithered exposures, each lasting 2624~s, eventually combined to a single reduced data cube. The details of the observations and their reduction can be found in \citet{Comerón2019}. The MUSE field-of-view of IC 1553 includes one known neighbouring galaxy LEDA 776941 (marked in red in Figure~\ref{Fig: color}) as well as a foreground star. We utilized MUSE white light images of IC 1533 to create a mask \& eliminate these known interloping objects. 
%______________________________________%

\section{MUSE data Analysis}
\label{sect: analysis}

\subsection{Stellar population analysis}
\label{sect: star_analysis}

We applied the methods outlined in \citetalias{Somawanshi} to extract the stellar kinematics and populations for IC~1553, with brief specific details provided here. For a full description, refer to section 3 of \citetalias{Somawanshi}. Voronoi-binning \citep{Cappellari03} was used to obtain spatially binned spectra for IC~1553. We adopted the same minimum S/N values per bin and per spaxel of 60 and 3, respectively, within the 4800–5500~\AA~range, as used for ESO~544-27. This resulted in a total of 1312 bins for IC~1553. This process was carried out using the GIST (Galaxy IFU Spectroscopy Tool) Pipeline \citep{BITTNER}. 

We fitted the spectra using the Penalized Pixel-Fitting (pPXF) code as outlined in \citet{Cappellari2004,Cappellari2017}, employing the semi-empirical (sMILES; \citealt{2023Knowles}) single-stellar-population (SSP) models, which match the sampling, resolution (FWHM), and wavelength range of MILES \citep{Vazdekis15}. The sMILES models were chosen for their higher resolution in [$\alpha$/Fe] abundances (ranging from -0.2 to +0.6 in steps of 0.2). However, we used the Kroupa IMF \citep{Kroupa01} for IC 1553 instead of the unimodal Salpeter IMF \citep{Salpeter55} that was used in \citetalias{Somawanshi}. While the choice of IMF has minimal impact on the mass fraction, as discussed in \citetalias{Somawanshi}, we find marginally lower fraction of unphysical fitted stellar populations (see \citetalias{Somawanshi} for details; also discussed briefly in next subsection) when using the Kroupa IMF.

For each spatial bin, nebular dust extinction was derived from the binned spectra through the Balmer decrement, measured using the H$\alpha$ and H$\beta$ emission line fluxes. This approach effectively identifies the presence and spatial distribution of dust within the galaxy. The nebular extinction values were then utilized to estimate the stellar extinction in the same spatial bin, following the relation outlined by \citet{Calzetti94}. The spatial distribution of $A_{v}$ values is presented in Figure~\ref{Fig: Mean_values}a. This is different from the approach of determining $A_{v}$ from the stellar continuum that was used in \citetalias{Somawanshi}, as the extinction from the dust lane in IC~1553 was not adequately accounted for using the $A_{v}$ determined by pPXF from the stellar continuum.

As in \citetalias{Somawanshi}, emission lines were masked to avoid any potential impact on the stellar population model fitting.
Stellar kinematics and populations were fitted separately using an additive \& multiplicative polynomial of 8th order. This was done with a regularization parameter of 0.07. As mentioned in \citetalias{Somawanshi}, this value was chosen to balance smoothing the solutions to reduce noise while preserving the star-formation history information and is similar to the approach followed by \citet{2019PinnaA} and \citet{2021Scott}. We obtained the mass weights for each SSP model using pPXF across all 1312 spatial bins of IC~1553.

%______________________________________

\subsection{Mass-weighted mean stellar population properties}
\label{sect: mean_maps}

From the mass weights, we excluded contributions from metal-poor ($-1.79\leq$[M/H]$\leq-0.96$) and low-alpha ($-0.2\leq[\alpha/\text{Fe}]\leq0.0$) stellar populations, which were marked as non-physical in \citetalias{Somawanshi} due to the limited capability of pPXF to accurately fit stellar population models, especially in regions affected by dust attenuation. Such a metal-poor and lower [$\alpha/\text{Fe}$] stellar population is not predicted for
any galaxy, neither in massive galaxies (e.g. \citealt{Vincenzo18}) nor for local group dwarf galaxies (e.g. \citealt{Kobayashi20b}). For a more detailed discussion of this unphysical stellar population, see \citetalias{Somawanshi}.  Subsequently, the remaining mass weights were utilized to determine the mean age, [M/H], and [$\alpha$/Fe] for each spectral bin in IC~1553. Their spatial distribution is illustrated in Figure~\ref{Fig: Mean_values}. We note that the mean age map is not very reliable due to tendencies of pPXF to favour majority older (>9~Gyr old) stellar population model fitting solutions balanced with small fractions of younger ($\sim$1--2~Gyr old) stellar populations \citep{Woo24, Wang24}.

The <[M/H]> distribution (Figure~\ref{Fig: Mean_values}c), in particular, reveals that the bulk of the stellar population is metal-rich (<[M/H]> $\geq 0.06$) but a region on the disc plane (outlined in black in Figure~\ref{Fig: Mean_values}c) stands out as being metal poor (<[M/H]> $\leq -0.66$). We note that this is despite the tendies of pPXF to favour relatively metal-rich ([M/H]$\sim$0) stellar population model fits \citep{Woo24}. 

This region is also found to have relatively higher [$\alpha$/Fe] compared to the mid-plane of IC~1553 (see Figure~\ref{Fig: Mean_values}d), though it doesn't stand-out in <age> (see Figure~\ref{Fig: Mean_values}b). There are additional bins adjacent to this region (within the dashed black ellipse marked in Figure~\ref{Fig: Mean_values}) that are also similarly metal-poor (see Figure~\ref{Fig: Mean_values}c), having higher [$\alpha$/Fe] than the mid-plane (see Figure~\ref{Fig: Mean_values}d) with a seemingly younger <age> (see Figure~\ref{Fig: Mean_values}b). 

Particularly the <[M/H]> and <[$\alpha$/Fe]> spatial distributions of IC~1553 thus reveal a region consisting of a few spatial bins that stand out in their stellar population chemistry compared to the rest of the galaxy. We select the 34 spatial bins with <[M/H]> $\leq -0.45$ and <[$\alpha$/Fe]> $\geq 0.25$ in this region as marked by a black outline in Figure~\ref{Fig: Mean_values}. 

Figure~\ref{Fig: Mean_values}a shows that the spatial bins within this region as well as the ellipse also exhibit significantly lower extinction values compared to the bins in the rest of the midplane. This potentially indicates a foreground dwarf galaxy, with stellar population properties that are distinct from IC~1553, whose light dominates the observed spectral bins within the outlined region in Figure~\ref{Fig: Mean_values} while also contributing to some of the spectral bins within the ellipse region in Figure~\ref{Fig: Mean_values}. 

Differences between the stellar populations inhabiting the mid-plane of IC~1553 and those at greater scale heights are also evident from Figure~\ref{Fig: Mean_values}, indicating differences between the thin and thick discs of IC~1553. This will be addressed along with our full sample of star-forming edge-on discs studied with MUSE IFU observations in a future publication (Somawanshi et al. in prep).

\begin{figure}
    \centering
    
    \includegraphics[width=\linewidth]{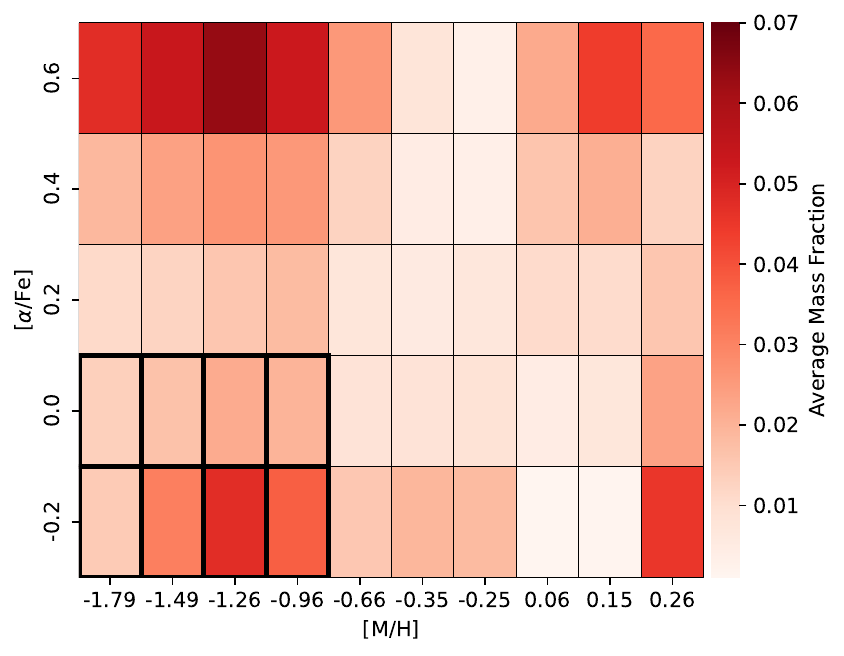}
    \includegraphics[width=\linewidth]{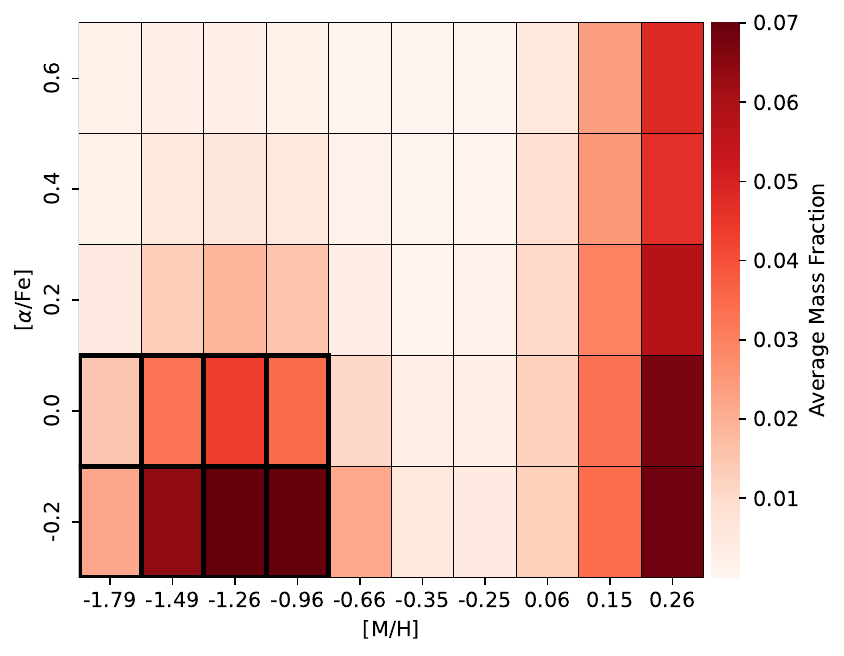}

    \caption{The [$\alpha$/Fe] vs [M/H] plane for stellar populations of all ages in the potential dwarf region [top] as well as in the remaining bins in IC~1553 [bottom]. The plane is binned as per the [$\alpha$/Fe] and [M/H] grid described in Section~\ref{sect: analysis} with each grid region coloured by the determined average mass fraction. The non-physical stellar population is outlined with thick black lines.}
    \label{Fig: population}
    
\end{figure}

\begin{figure*}
    \includegraphics[width=\textwidth]{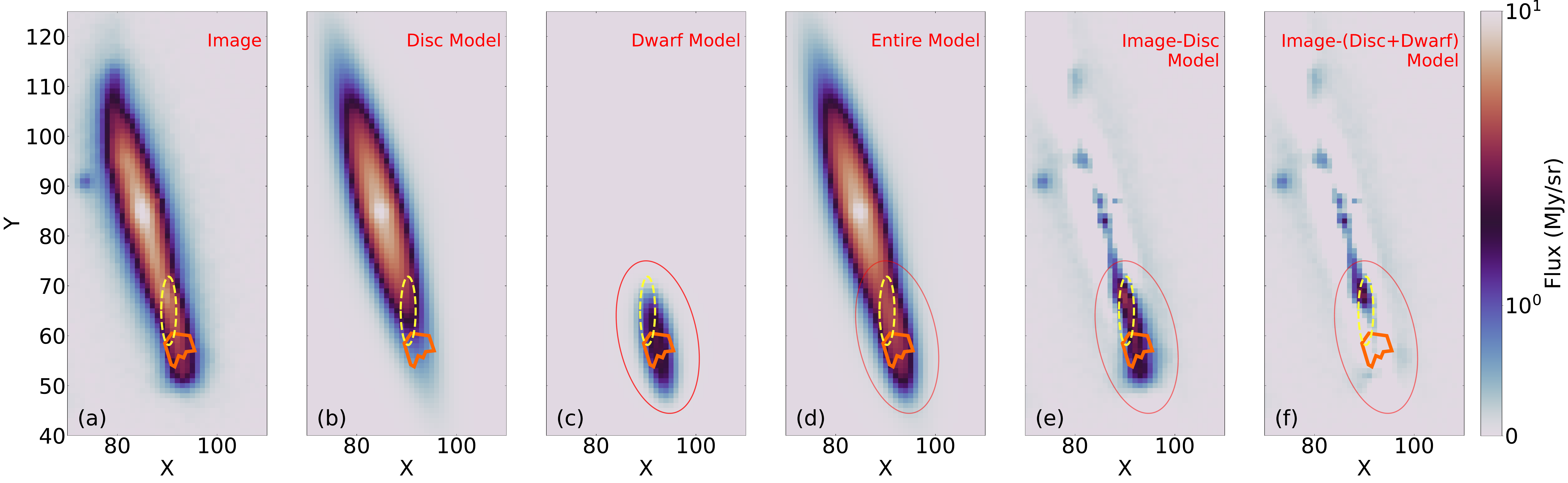}

    \caption{(a) The S$^{4}$G survey \textit{Spitzer} IRAC 3.6~$\mu m$ image of IC~1553 is shown along with GALFIT modeled (b) disc galaxy, (c) dwarf galaxy and (d) entire model (disc+dwarf). Also shown are residuals produced when subtracting from the image (e) just the modeled disc galaxy and (f) the entire galaxy model. All panels show also the analysed dwarf region (orange outline) and region of interest (yellow dashed ellipse) from MUSE as in Figure~\ref{Fig: Mean_values}. The elliptical aperture used to compute the dwarf galaxy mass is also marked in red in panels (c)--(f).}
    \label{Fig:Imaging_data}
\end{figure*}

\subsection{[$\alpha$/Fe] vs [M/H] in the dwarf region of the galaxy} 
\label{sect:dwarf_alpha}

Figure~\ref{Fig: population} [top] illustrates the average mass fraction across all spatial bins of the potential dwarf region in the [$\alpha$/Fe] vs [M/H] plane for stellar populations of all ages. The region contains a significant metal-rich stellar population with [M/H] $\geq 0.15$ and [$\alpha$/Fe] $= 0.6$, alongside a stellar population characterized by [M/H] = 0.26 and [$\alpha$/Fe] $= -0.2$.  
These stellar populations are similar to those of ESO~544-27 discussed \citetalias{Somawanshi} where the higher and lower [$\alpha$/Fe] stellar populations were considered being associated with its thin and thick disc. This is likely also the case here and will be discussed further considering also the region beyond the dwarf region in Somawanshi et al. (in prep.). A metal-poor and lower [$\alpha$/Fe] stellar population (-1.79 $\leq$~[M/H]~$\leq$-0.96 and -0.2 $\leq$~[$\alpha$/Fe]~$\leq$ 0.0) is also present (as outlined) in Figure~\ref{Fig: population} but should be ignored as already discussed in Section~\ref{sect: mean_maps}.

Of particular interest in Figure~\ref{Fig: population} [top] is that we observe a metal-poor higher [$\alpha$/Fe] stellar population with [$\alpha$/Fe] $= 0.6$ and -1.7 $\leq$~[M/H]~$\leq$ -0.96 in this region. With a mass fraction of $\sim 0.217 \pm 0.006$, this metal-poor higher [$\alpha$/Fe] population thus forms about a fifth of the total stellar mass in the marked area in Figure~\ref{Fig: Mean_values}. In comparison with Figure~\ref{Fig: population} [bottom], which shows the average mass fraction in spatial bins beyond the dwarf region, it is evident that this metal-poor higher [$\alpha$/Fe] population belongs exclusively to the potential foreground dwarf galaxy and is not present in other parts of the galaxy. However, the metal-rich lower and higher [$\alpha$/Fe] stellar populations likely have contributions from both IC~1553 and the potential foreground dwarf.

The dwarf galaxy is thus identified with higher [$\alpha$/Fe] and more metal-poor stellar populations relative to the disc of IC 1553. We note that for this population, pPXF gives the highest value of the population grid ([$\alpha$/Fe] = 0.6), while lowering metallicity ([M/H]=-1.79 to -0.96) and maximizing age ($\sim$12~Gyr). Similar tendency is found for age and metallicity \citep[e.g.][]{Sattler24}, tested extensively with mock galaxy spectra by \citet{Woo24} and \citet{Wang24}. Therefore, to determine the exact value of [$\alpha$/Fe], more detailed analysis would be required

%We note here that pPXF has a tendency to favour extrema solutions for stellar populations of star-forming galaxies, i.e., when the integrated stellar light consists of a physical juxtaposition of light from mostly higher-$\alpha$ metal poor stars from the dwarf and metal rich stars from the edge-on disc (both lower and higher-$\alpha$), pPXF can identify the distinct stellar populations from the integrated light but tends to favour solutions from the edges of the model grids (as seen from Figure~\ref{Fig: population}). While most previous studies utilised the MILES model grid with fixed [$\alpha$/Fe] values, the favoring of extrema solutions was also noted in the age-metallicity grid by \citet[see e.g.]{Sattler24}. Extensive tests by \citet{Woo24} and \citet{Wang24} from mock spectra also showcase the tendencies of pPXF to favour majority older (>9~Gyr old) stellar population model fitting solutions balanced with small fractions of younger ($\sim$1--2~Gyr old) stellar populations.

The presence of the metal-poor higher [$\alpha$/Fe] population spread over multiple adjoining spectral bins in a specific spatial location, may imply the possibility of a foreground dwarf galaxy, but we need to corroborate its identification in the stellar population parameter space with potential kinematic and photometric signatures.

\subsection{Dwarf kinematics}
\label{sect: dwarf_old} 

It is important to assess whether a distinct dwarf galaxy can be discerned in this region in the kinematic spatial distribution of IC~1553. Figures~\ref{Fig: Mean_values}e \& ~\ref{Fig: Mean_values}f respectively present the mean stellar velocity (V) and stellar velocity dispersion ($\sigma$), as derived from the GIST pipeline, as described in Section~\ref{sect: star_analysis}. Figure~\ref{Fig: Mean_values}e shows that the dwarf region is moving in the direction of rotation of the disc galaxy. 
Additionally, Figure~\ref{Fig: Mean_values}f shows that the spatial bins within the dwarf region have a lower mean stellar velocity dispersion ($\sigma\rm_{dwarf}= 21.6 \pm 2.7$) compared to the spectral bins of IC~1553 diametrically opposite to the dwarf region with $\sigma \sim 51$. 
Likewise, the bins within the ellipse region also exhibit similarly low stellar velocity dispersion values. If the dwarf were actively disrupting the disc of IC~1553, we would expect to observe a higher velocity dispersion in the dwarf region. Conversely, if it were physically too distant from IC~1553 albeit within the line-of-sight, the stellar velocity would differ significantly from the galaxy's velocity field resulting in a higher measured mean velocity dispersion in the dwarf region. 

The relatively lower velocity dispersion suggests that the dwarf is likely a dynamically colder satellite near the disc of IC~1553 that lies in the foreground in the line-of-sight. The dwarf itself may still be disrupting with an extended star-forming tail in the ellipse region marked in Figure~\ref{Fig: Mean_values} but it doesn't seem to be affecting the disc of IC~1553.

% The standard deviation of the stellar velocity for the 34 bins in the dwarf region of IC~1553 is approximately $\sim9.469$, whereas for the rest of the galaxy, it is significantly higher at $\sim70.919$. Additionally, the mean stellar velocity dispersion calculated from figure \ref{Fig: Mean_values} [sixth] for the dwarf spatial bins is $\sim4.561$, and for the rest of the galaxy mean value is $\sim1.248$. Meanwhile, the mean stellar velocity of the dwarf region from figure \ref{Fig: Mean_values} [fifth] is $\sim-114.350$. 

%___________________________________

\section{\textit{Spitzer} Imaging Data Analysis}
\label{sect: Imaging_analysis}

\subsection{Presence of the dwarf galaxy in photometry}
\label{sect: Dwarf_Detection}

Utilising the S$^{4}$G survey \textit{Spitzer} IRAC 3.6~$\mu m$ image of IC~1553 (Figure~\ref{Fig:Imaging_data}a) and considering also that its classified as an edge-on disc with a bright nuclear point source \citep{Salo15}, we make a suitable GALFIT \citep{Peng10} model (Figure~\ref{Fig:Imaging_data}b) with a point spread function and a sérsic profile (magnitude: $13.13$, sérsic index: 0.64, half-light radius: 16.55 pixels). The residual image (Figure~\ref{Fig:Imaging_data}e), from subtracting the modeled disc galaxy from the observed image, clearly shows an extended feature. The dwarf region and the extended region of interest identified from the stellar population analysis with MUSE (see Figure~\ref{Fig: Mean_values}) is consistent in position with this extended feature. When the dwarf galaxy is modeled separately (Figure~\ref{Fig:Imaging_data}c; centered roughly at the location of the dwarf region from Figure~\ref{Fig: Mean_values}) using a single sérsic profile in GALFIT (magnitude: $15.49$, sérsic index: 0.17, half-light radius: 7.19 pixels) and included to model the entire image (Figure~\ref{Fig:Imaging_data}d), the extended feature within the elliptical outline is significantly reduced in the resulting residual when this entire model in subtracted from the image (Figure~\ref{Fig:Imaging_data}f), though some residual flux towards the center of the edge-on disc remains in the dwarf region. 

We note that the dwarf galaxy appears more extended in the model than in the MUSE region (Figure~\ref{Fig:Imaging_data}c) as the stellar populations of the dwarf galaxy only dominate that of the background bright disc galaxy within that smaller region. The outer fainter regions of the dwarf galaxy are not bright enough relative to the background disc galaxy to impact the mean stellar population properties in these outer regions.

\subsection{Mass of the dwarf galaxy}
\label{sect: Mass_estimate}

To measure the mass of the dwarf galaxy, and its relative mass fraction compared to the modeled disc galaxy, we consider an elliptical aperture (marked by red in Figure~\ref{Fig:Imaging_data}) centered on the dwarf galaxy with the same position angle and ellipticity, but twice the effective radius, thus encompassing the entire dwarf galaxy region. In this aperture, we measured the flux from the dwarf model as $\sim 172.328$~MJy/sr. To quantify the goodness of our model, we compute the ratio of the total fluxes, within the elliptical aperture, of the residual image (Figure~\ref{Fig:Imaging_data}f) and the original image (Figure~\ref{Fig:Imaging_data}a) as $7.9\%$, demonstrating that our model successfully captures the majority of the galaxy’s light. We then utilized Equation 6 from \citet{2015Querejeta} to estimate the mass of the dwarf galaxy which converts 3.6 $\mu m$ old stellar flux into stellar mass with an accuracy of $\sim 0.1$ dex \citep{2014Meidt}. As per their suggested mass-to-light ratio, $M/L=0.6$, and assuming a Chabrier IMF \citep{Chabrier03}, we obtain the mass of the dwarf galaxy to be $\sim 1.27 \pm 0.31 \times 10^{9} \rm~M_{\odot}$. We then applied the same elliptical aperture to the entire model to measure the total flux contribution of IC~1553 and the dwarf in the region of interest. This comes out to be $\sim 447.82$ MJy/sr, resulting in the computed mass of this entire region to be $\sim 3.33 \pm 0.71 \times 10^{9} \rm~M_{\odot}$. Thus, the modeled dwarf galaxy has a mass fraction of $0.39 \pm 0.18$ within the elliptical aperture in Figure~\ref{Fig:Imaging_data}. 

This value is higher, though within error, than the mass fraction of $\sim 0.22 \pm 0.01$ for the metal-poor, higher-[$\alpha$/Fe] population in the dwarf from MUSE (see Figure~\ref{Fig: population} and Section~\ref{sect:dwarf_alpha}). While the former is the ratio of the mass of all stellar populations in the dwarf galaxy to that of all stellar populations in the edge-on disc, the latter is the ratio of the mass of the higher-[$\alpha$/Fe] metal-poor population, that is only present in the dwarf galaxy, to that of all other stellar populations both in the dwarf and the background IC 1553. It thus follows that the mass fraction of the metal-poor, higher-[$\alpha$/Fe] stellar population in the dwarf galaxy (computed from MUSE IFU stellar population model fitting) should be lower than the mass fraction of the dwarf galaxy (computed from \textit{Spitzer} image modeling). The two mass fractions are thus not inconsistent. 

We note here that the mass fraction computed from the \textit{Spitzer} image modeling is an approximate estimate that enabled comparison with the MUSE IFU stellar population mass fraction. A more accurate mass estimate would require modeling of the disrupting dwarf galaxy, which is beyond the scope of this work. 

%___________________________________

\section{Summary and conclusion}
\label{sect: discussion_and_conclusion}

We thus report the identification of a star-forming dwarf galaxy, having mass $\sim 1.27 \pm 0.31 \times 10^{9} \rm~M_{\odot}$, that is co-spatial with the luminous edge-on disc galaxy IC~1553. The dwarf galaxy is initially identified from stellar population model fitting of MUSE IFU integrated light observations (Figure~\ref{Fig: Mean_values}c,d) due to its higher [$\alpha$/Fe], metal-poor stellar population ([$\alpha$/Fe] = 0.6 and  -1.79 $\leq$~[M/H]~$\leq$-0.96; see Section~\ref{sect:dwarf_alpha}), in contrast to the metal-rich stellar populations with lower mean [$\alpha$/Fe] that inhabit the disc of IC~1553. In this work, we have thus mainly utilized the stellar population model fitting with pPXF to chemically distinguish the dwarf galaxy from the background edge-on disc, though additional analysis may be required to more accurately obtain the [$\alpha$/Fe] and [M/H] of the dwarf galaxy. 

This identification is consistent with how the higher [$\alpha$/Fe] and metal-poor stellar populations identified in dwarf spheroidal satellites of MW and M31 \citep[e.g.][]{2011Kirby,2020Kirby} contrast the lower [$\alpha$/Fe] and metal-rich stellar populations of the MW and M31 thin discs \citep[e.g.][]{Hayden15,Arnaboldi22,Kobayashi23}. 

This dwarf galaxy is consistent with being a satellite of IC~1553 as its velocity is not too dissimilar from that of the rotating disc of IC~1553 (which would have been the case if it were a foreground galaxy), but its dynamically colder than the background edge-on disc (Figure~\ref{Fig: Mean_values}e,f). The dwarf galaxy is further identified from the GALFIT modeling of the \textit{Spitzer} IRAC 3.6~$\mu m$ image of IC~1553 (Section~\ref{sect: Imaging_analysis}). 

The presence of some residuals upon consideration of the dwarf galaxy in the GALFIT modeling (Figure~\ref{Fig:Imaging_data}f), as well as extended features near the position of the dwarf galaxy in the stellar population maps obtained from MUSE IFU images (dotted ellipse in Figure~\ref{Fig: Mean_values}), indicates that the dwarf is disrupting with an extended star-forming tail. The stars in the tail likely have a higher <[M/H]> (Figure~\ref{Fig: Mean_values}c) but similar <[$\alpha$/Fe]> (Figure~\ref{Fig: Mean_values}d) as the main body of the dwarf galaxy.

Earlier studies \citep{Sattler24} where stellar population models for MUSE IFU observations of IC~1553 were fitted only in the age and [M/H] parameter space, did not enable the co-spatial dwarf galaxy to be identified. Our methodology \citepalias{Somawanshi} with the expanded parameter space including [$\alpha$/Fe] in the sMILES stellar population models allowed the reliable identification of the dwarf satellite galaxy in <[$\alpha$/Fe]> and <[M/H]> maps (Figure~\ref{Fig: Mean_values}c,d). This is the first such dwarf galaxy to be identified from its chemical signature in integrated light IFU observations.
 
%___________________________________

\section*{Acknowledgements}
We thank the anonymous referee for their comments.
SB was funded by the INSPIRE Faculty Award (DST/INSPIRE/04/2020/002224), Department of Science and Technology (DST), Government of India. The award also supported DS during his recurring stays at IUCAA, Pune, India. PKM is supported by the KIAS Individual Grant (PG096701) at the Korea Institute for Advanced Study. We acknowledge the use of Pegasus, the
high-performance computing facility at IUCAA. This research made use of Astropy-- a community-developed core Python package for Astronomy \citep{Rob13}, SciPy \citep{scipy}, NumPy \citep{numpy} and Matplotlib \citep{matplotlib}. This research also made use of NASA’s Astrophysics Data System (ADS\footnote{\url{https://ui.adsabs.harvard.edu}}).

%%%%%%%%%%%%%%%%%%%%%%%%%%%%%%%%%%%%%%%%%%%%%%%%%%
\section*{Data Availability}
The reduced MUSE data cube for ESO IC 1553 is publicly available through the ESO Phase 3 data release at \url{https://doi.eso.org/10.18727/archive/8}. Meanwhile, the imaging data and PSF for IC~1553 are publicly available through S$^{4}$G data release at \url{https://irsa.ipac.caltech.edu/data/SPITZER/S4G/overview.html}.


\bibliographystyle{mnras}
\bibliography{example} 
%%%%%%%%%%%%%%%%%%%%%%%%%%%%%%%%%%%%%%%%%%%%%%%%%%

% Don't change these lines
\bsp	% typesetting comment
\label{lastpage}
\end{document}